# Biodegradable, Millimeter-Scale Light-Emitting Sensors for Distributed Environmental Monitoring – Functional Pixie Dust

Zhiming Hu[1,2#], Danzhen Zhang[1#], Janghun Ko[3#], Haohui Zhang[4#], Jiale Chen[1,5#], Chanho Park[6,7#], Jiatong Zhang[1,5], Qiuna Zhuang[1], Shiwei Xu[1], Xiaoran Yang[8], Dain Son[3], Taehoon Kim[1], Uikang Joo[3], Zhaojian Xu[1], Hyunsoo Kim[3], Richard Chai[9], Gwangmin Bae[1], Wooyoul Maeng[1], Qiong Wang[1], Sangmin Lim[3], Liangsong Zeng[1,8], Un-Seong Baik[3], Kaiqing Zhang[4], Liming Yuan[1,5], Yonggang Huang[4,5,8,10*], Jin-Tae Kim[3*], John A. Rogers[1,5,8,10,11*]

[1]Center for Bio-Integrated Electronics, Northwestern University, Evanston, IL, USA.

[2]School of Advanced Manufacturing Engineering, Nanjing University, Suzhou, China.

[3]Department of Mechanical Engineering, Pohang University of Science and Technology, Pohang, Republic of Korea.

[4]Department of Civil and Environmental Engineering, Northwestern University, Evanston, IL, USA.

[5]Department of Materials Science and Engineering, Northwestern University, Evanston, IL, USA.

[6]Department of Chemistry, Sejong University, Seoul, Republic of Korea.

[7]Future Chemical Materials Research Institute, Sejong University, Seoul, Republic of Korea.

[8]Department of Mechanical Engineering, Northwestern University, Evanston, IL, USA.

[9]Department of Chemistry, Northwestern University, Evanston, IL, USA.

[10]Querrey Simpson Institute for Bioelectronics, Northwestern University, Evanston, IL, USA.

[11]Department of Biomedical Engineering, Northwestern University, Evanston, IL, USA.

[#]These authors contributed equally: Zhiming Hu, Danzhen Zhang, Janghun Ko, Haohui Zhang, Jiale Chen, Chanho Park.

Corresponding author. Email: y-huang@northwestern.edu; jimmy516@postech.ac.kr; jrogers@northwestern.edu

## Abstract

Methods for large-area, precise monitoring across natural environments are of growing interest due to pressing needs for sustainable management of rapidly increasing anthropogenic activities. Established approaches involve sparse spatial sampling and/or sequential measurements, while emerging techniques exploit miniaturized electronics or passive optical methods. Various constraints in scalability, costs, robustness, operational range and other factors create a need for alternatives. Here, we introduce a concept that overcomes many of these limitations through the combined use of chemically induced light emission and chemically responsive optical filter elements in millimeter-scale systems that we refer to as functional pixie dust (fPD) sensors, designed specifically for monitoring natural water systems during nighttime to eliminate background optical interference and to enhance remote analysis. These floating devices act as Lagrangian tracers to follow surface flows and to simultaneously measure the concentrations of key chemical species along their trajectories. Optimized designs exploit environmentally compatible constituent materials that are also degradable through natural processes to benign end products, thereby eliminating the need for recovery. Spatially and spectrally resolved ratiometric measurement schemes ensure robust operation and ability to address practical requirements in range, operational lifetime, time response and sensitivity. Demonstrations include distributed measurements of pH, $Hg^{2+}$, and $NO_2^-$, each of relevance to industrial discharge, toxic metal contamination, and nitrogen-rich runoff, adapted for static concentration gradients, flow-driven transport conditions, and outdoor aquatic settings. The results establish a framework for environmental sensing using degradable, self-powered microsystems capable of scalable deployment and remote readout.

## Introduction

Large-area, real-time monitoring of key physical and chemical parameters in complex environments remains a central challenge for sustainable ecological management and early warning of hazardous conditions, of growing importance as anthropogenic and natural hazards become increasingly intertwined[1-4]. Existing mobile and fixed monitoring systems rely on sequential scanning and sparse spatial coverage, limiting their ability to provide continuous, high-resolution and multi-parameter sensing in dynamic environments[5-8]. An alternative approach involves large numbers of remotely addressable sensing units distributed across an area of interest[9,10]. In lightweight miniaturized designs, these devices can move with ambient air and water flows without perturbation, thereby enabling continuous measurements of spatio-temporal processes in a Lagrangian frame[11-14]. Optimized structures can enable broad dispersal and stable transport through these aerodynamic or hydrodynamic interactions, in some cases using wind- and water-dispersed seeds as biological inspiration[15-18]. Engineered microstructures with carefully controlled fluid dynamic behavior may serve as passively dispersed sensing nodes configured with wireless readout capabilities[19,20]. For monitoring of natural water systems, patterns of flow along with measurements of acidification and alkalinization, industrial metal contamination, and nutrient-related eutrophication, together with their spatio-temporal spread, represent assessment metrics that could be captured in this manner to guide remediation or rescue efforts[21-24].

A key requirement for systems of these types is environmental compatibility and degradability[25]. Specifically, after a defined operational lifetime, the devices should ideally disappear naturally into benign byproducts to avoid the need for large-scale recovery and disposal, which would be impractical for envisioned wide-area deployments. Platforms that use advanced electronic and optoelectronic components offer attractive sensing capabilities, but may involve prohibitive costs and system complexity for implementation at scale[26-28]. Colorimetric approaches that rely on reflection of ambient light offer simple solutions, although effective use at a distance in complex outdoor environments remains challenging, particularly in the presence of bright daylight or during low-light conditions. Alternative passive optical strategies that exploit fluorimetric schemes offer some advantages, but require high-power, short-wavelength scanning illumination to generate measurable levels of emission[14].

The mechanisms of chemiluminescence (CL) provide an option for a self-powered light source that can be integrated with a colorimetric sensing layer that serves as an optical filter to convert chemical information into stable color-encoded optical signals suitable for scalable remote sensing, with background-free readout when evaluated in the dark. The following introduces engineering principles for devices that realize this concept in materials that are entirely benign and environmentally degradable, with a focus on self-orienting floating platforms that faithfully follow surface flows and on optical architectures that enable efficient remote readout of aqueous chemical species of interest. The resulting designs allow for spatially and spectrally resolved ratiometric optical analysis, thermally enhanced CL emission, and optimized light management. Comprehensive experimental and theoretical studies establish the underlying materials chemistry aspects, optical engineering features, thermal characteristics and hydrodynamic properties. Demonstrated capabilities span quantitative spatial and temporal sensing of environmentally relevant analytes in natural water systems, including pH, heavy metal concentrations, and nitrite levels, of concern with industrial discharge and agricultural activity.

## Results

### Materials and Engineering Designs

In the simplest forms, the structures introduced here represent light-emitting Lagrangian tracers optimized for monitoring flows in low light conditions or at nighttime, referred to as pixie dust (PD) devices. The emphasis here is on functional versions, i.e. fPD systems, that can operate in both passive and active modes for multi-modal, remote monitoring of natural water systems during both daytime and nighttime (Fig. 1a). Remote digital imaging of these devices yields temporally and spatially resolved chemical information for reconstruction of dynamically evolving concentration gradients and chemical maps of multiple species of relevance to sustainable management (Fig. 1b).

To facilitate measurements in natural water systems, the structural designs incorporate simple buoyant structures like those found in lotus seeds (Fig. 1c,i). (Other bio-inspired layouts are also possible, including those in geometries of wind-dispersed seeds (*Tristellateia australasiae*; Extended Data Fig. 1a-c) and lily pads (Extended Data Fig. 1d-f).) These floating devices function in a Lagrangian frame[8] of reference, drifting with the local water flow and measuring environmental conditions along their trajectories. The fPD devices described here include, from top to bottom, (1) a cylindrical shell of silk fibroin (1.5 mm diameter, 1.5 mm length, 0.1 mm thickness) to form an air cavity that ensures sufficient buoyancy, (2) an agarose hydrogel (0.5 mm thickness, 4 mm outer diameter and 1.5 mm inner diameter ), doped with colorimetric chemical reagents to yield an optical filter, referred to in the following as a colorimetric filter, that responds to species of interest, (3) a film of silk fibroin (0.3 mm thickness and 4.5 mm diameter) to isolate the top and bottom parts of the device, (4) a cylindrical volume (0.6 mm depth and 1.5 mm diameter) of an aqueous solution of a fluorescent species placed in the central cavity (per 7) to serve as an (optional) source of light for ratiometric analysis, (5) a ring-shaped volume (3 mm outer diameter, 1.6 mm inner diameter, 0.6 mm depth) of CL reactants placed in the surrounding annular cavity (per 7) to produce light emission, (6) a solid ring structure (3 mm outer diameter, 1.6 mm inner diameter, 0.1 mm thickness) containing a chemistry confined within the surrounding annular cavity (per 7) that reacts in an exothermic process with water to increase the intensity of CL emission, (7) a concentric annular reservoir (outer ring: 4.5 mm outer diameter, 3 mm inner diameter; inner well: 1.5 mm outer diameter, 1.1 mm inner diameter; depth: 0.6 mm) of silk fibroin to house the CL and fluorophore solutions and (8) a reflector formed from a layer of magnesium (~1 μm thickness) on the sidewalls and base of the reservoir to ensure that light only emerges from the device by passage through the colorimetric filter, and (9) a conical mass tuner (4.5 mm diameter, 2 mm height) based on porous silk fibroin to regulate buoyancy and stabilize device orientation. This architecture enables robust sensing during nighttime, via combined CL emission, colorimetric filtering and remote digital image analysis. Daytime operation is possible with or without CL emission. The following focuses on the active, light-emitting mode.

Interactions between chemical reagents within the colorimetric filter and targeted species that diffuse into this layer from the surrounding water induce changes in its optical absorption properties. Corresponding modifications of the spectral content of light generated by CL that passes through this layer correlate quantitatively with the concentration of that species (Fig. 1d). In one operating scheme, spatially separated sensing and reference regions defined by the annular ring geometry of the colorimetric filter enable spatially resolved ratiometric analysis. To illustrate this scheme, top-down images acquired during the daytime and nighttime reveal pronounced concentration-dependent color variations in the sensing region, while the reference region remains unchanged. Remote evaluation can be performed using digital cameras mounted on aerial drones or on fixed structures. The overall designs can extend to dimensions in the millimeter range (~3 mm for the example presented here), enabling cost-effective deployment at scale and over large areas. Corresponding daytime and nighttime photographs show the devices operating in passive reflective and active emissive modes, the latter based on red, green, and blue CL emission (Fig. 1e-h, Extended Data Fig. 2). Top-view images of a representative device placed next to a lotus seed provide a size comparison, emphasizing the bio-inspired design (Fig. 1i,j). Side-view images of a device without the reflector reveal the multilayer architecture and overall buoyant design (Fig. 1k,l). These pictures also highlight the top air cavity and conical base to promote self-orientation in aqueous environments.

As mentioned previously, the devices rely on non-hazardous materials that degrade harmlessly in the environment, thereby avoiding the need for recovery and disposal after a period of use. Accelerated degradation studies in Ajisawa's reagent at 70 °C demonstrate progressive structural breakdown and chemical degradation over the course of two weeks (Fig. 1m-p). The non-functional materials, from silk to magnesium, degrade according to chemical processes described elsewhere[29,30]. Details associated with the functional materials appear in a subsequent section. As degradation proceeds, the magnesium reflector dissolves within the first day, the microchannel architecture gradually collapses, and the conical base deforms as the air cavity and supporting framework dissolve. Carefully selected materials and structural designs ensure reliable operation throughout the practical nighttime operating period of approximately 15 min, limited primarily by the decrease in CL emission intensity and the sensitivity of the optical readout system, prior to degradation.

## Hydrodynamic, optical and thermal characteristics

Engineered distributions of mass elements and buoyancy structures ensure stable hydrodynamic behavior. Specifically, the center of gravity (CG) of the basic PD structures described above resides approximately 0.03 mm below the center of buoyancy (CB), thereby generating a restoring moment and maintaining a stable orientation under any angular perturbation (Fig. 2a). Upon submersion, this offset drives rapid self-orientation and ascent to the surface regardless of its initial orientation (Fig. 2b). Time-resolved particle image velocimetry (TR-PIV) reveals that transient wake structures formed during reorientation reorganize into symmetric vortices during steady ascent, consistent with stable and balanced motion (Extended Data Fig. 3a). Upon reaching the water surface, buoyancy, gravity and capillary forces balance to establish a partially submerged equilibrium configuration. This floating state maintains continuous contact between the colorimetric filter and the surrounding water while preserving an upward-facing optical readout surface, enabling simultaneous chemical sensing and remote imaging. In agreement with analytical modelling and theoretical estimates (Supplementary Notes 1-3), quantitative tracking shows that the PD devices reach equilibrium orientation (Fig. 2c) and terminal velocity within ~3 s (Fig. 2d). The resulting trajectories exhibit nearly vertical ascent with only slight lateral dispersion (Extended Data Fig. 3b). Water-tunnel experiments further validate the ability of PD devices to function as Lagrangian flow tracers. The trajectories follow streamlines around bluff bodies (Fig. 2e, Extended Data Fig. 3c), with velocity profiles that agree closely with TR-PIV measurements (Fig. 2f, Extended Data Fig. 3d), thus demonstrating the ability to reconstruct flow fields and to capture inherently Lagrangian characteristics, such as dispersion (Supplementary Fig. 3).

For fPD sensors, operation in the active mode requires an optical design that efficiently channels light from CL emission, and optionally from fluorescence, through the annular colorimetric filter structure. Monte Carlo simulations quantify the key effects (Fig. 2g,h, Supplementary Note 4). For instance, the reflector directs photons toward the top part of the device, thereby increasing the detectable emission that passes through the filter structure by ~40%, consistent with experiment (30-50%; Extended Data Fig. 4a,b). As a representative case for pH sensing, the simulated output agrees with the measured trends for devices with and without the reflector (Extended Data Fig. 4c,d).

The overall emission power, and thus the remote-sensing range for a given imaging system, can be enhanced by accelerating the rate of the CL reaction by increasing the temperature (Extended Data Fig. 4e,f). Exothermic reactions between CaO, as the solid ring structure referenced previously, and water in the CL reservoir provide thermal power for this purpose, with a product (calcium hydroxide) that is environmentally compatible. Finite element modeling (Fig. 2h) indicates an increase of ~18 °C within the first 2 s for the design outlined in Fig. 1. The highest temperature occurs near the central region of the device, consistent with experiment (Extended Data Fig. 5a,b). The emission power increases by a factor of approximately 1.6 with every increase in temperature of 10 °C (Fig. 2i). Designs of the CaO structure and the CL reservoir can be tailored to meet requirements that balance detection range, defined by emission intensity, with operational lifetime, defined by reaction duration (Extended Data Fig. 5c).

## Chemiluminescent emission and colorimetric sensing

CL emission originates from a peroxyoxalate reaction between divanillyl oxalate (DVO) and hydrogen peroxide. Oxalate decomposition generates a high-energy intermediate that transfers excitation energy to a fluorescent dye (Fig. 3a), with vanillin and carbon dioxide as final products. Dyes used in biological imaging applications are attractive choices for a range of emission colors (Fig. 3c). Normalized spectral measurements confirm well-separated emission bands for representative red, green, and blue dyes (Nile Red, Coumarin 6, and 7-diethylamino-4-methylcoumarin; Fig. 3d, Supplementary Fig. 4). The CL emission from the bulk reaction remains detectable for more than 240 min at room temperature with a reactant volume of 3 mL (Fig. 3e, Extended Data Fig. 6b,c), although the emission intensity decreases continuously as the reaction proceeds. The CL reaction rates can be suppressed by reducing the temperature (Extended Data Fig. 6d-f) for stable cold storage, such that activation can be triggered by warming immediately prior to deployment, further enhanced as necessary by exothermic reaction with the CaO structure.

The colorimetric filter consists of responsive chemistries immobilized within an agarose matrix. Anthocyanin, dithizone and a Griess reagent system (sulfanilamide and N-(1-naphthyl)ethylenediamine) serve as the basis for pH, heavy metal ($Hg^{2+}$) and nitrite sensing, respectively (Fig. 3b,f–h). These chemistries undergo changes in their optical absorption properties by amounts that depend quantitatively on the concentrations of the corresponding analytes. For anthocyanin, the dominant absorption band shifts from ~525 nm in acidic to ~600 nm in basic solutions (Fig. 3i), accompanied by a gradual increase in absorbance corresponding to a color change from red to green and then to purple. For dithizone, increasing the concentration of $Hg^{2+}$ from zero to 100 ppm induces progressively broader absorption across the visible spectrum, particularly in the 450–600 nm range. This broadband absorption leads to an overall darkening toward a brownish color (Fig. 3j). For the diazotization-based chemistry, increasing the concentration of nitrite enhances the absorption between 400 and 600 nm, accompanied by a change in color from pale to deep red. The monotonic increase in absorbance near ~620 nm provides an ideal basis for quantitative analysis (Fig. 3k). Additional details are in the Methods.

The timescales for these changes to occur within the fPD device structure follow from the corresponding reaction rates and the timescales for diffusion of the targeted species into the colorimetric filter. For cases examined here, the latter, which depends on the thickness of the filter, is the limiting step. For a given concentration of colorimetric species, selection of this thickness balances both the response time and the magnitude of the change in spectral characteristics of transmitted light. The optical attenuation, and thus the magnitude of the filter response, scales exponentially with thickness, according to the Beer-Lambert law; the time response scales like the square of the thickness and exponentially with temperature. For representative examples in the following, the measured response times are in the range of tens of seconds, consistent with simulations of the diffusion processes (Extended Data Fig. 7, Supplementary Note 7).

**Quantitative sensing through spatially and spectrally resolved methods**

The fPD platform supports two complementary modes for quantitative evaluations: (1) spatially resolved analysis based either on CL emission from the sensing and reference regions of a single device or from separate devices designed for sensing and for reference, and (2) spectrally resolved analysis based on analyte-dependent emission for sensing and analyte-independent fluorescence as a reference. Both approaches compensate for the gradual decrease in overall CL intensity that occur naturally as the reaction proceeds. The spatially resolved analysis used a mixed orange-emitting CL formulation containing red and green dyes to cover the filter-relevant wavelength range (~500–650 nm). In contrast, the spectrally resolved measurements used a green-emitting CL formulation to provide CL emission near 520 nm and excite the Cd-based QD reference emission near 710 nm, allowing the two peaks to be clearly resolved. Time-dependent CL emission spectra show a continuous decrease in absolute intensity without change in the spectral profile over the effective measurement window (Supplementary Fig. 5).

The following illustrates these concepts in the case of pH sensing. Here, with increasing pH, the intensity of CL light that passes through the colorimetric filter decreases systematically for wavelengths near 620 nm. By contrast, the feature near 520 nm varies in a nonmonotonic fashion with pH, increasing at intermediate pH values and decreasing under strongly acidic or alkaline conditions (Fig. 4a, Supplementary Fig. 6). These behaviors are consistent with the calculated spectra obtained from the product of the reference CL emission and filter transmittance (Supplementary Fig. 7a). Spatially resolved analysis, based either on a single device or on multiple devices, compares reference CL emission to that modulated by passage through the colorimetric filter. The corresponding ratio of these intensities at a given wavelength evolves systematically with pH across the visible range (Supplementary Fig. 7d). The feature near 620 nm shows the strongest monotonic dependence (Supplementary Fig. 7g), as expected based on the filter properties (Fig. 3i). This approach works reliably when the imaging system can separate the sensing and reference regions of a single device, or when multiple separate devices designed for sensing and reference can be resolved but spaced by distances smaller than the length scales associated with spatial variations in pH.

An alternative that avoids these requirements exploits spectral resolution enabled by the addition of a fluorophore into the central region of the device with an emission wavelength that is unaffected by the colorimetric filter. A demonstration of this concept uses quantum dots (QDs) selected for their efficient excitation with green light from CL emission to yield fluorescence at a wavelength near 710 nm. This fluorescence provides an internal reference whose magnitude varies with CL emission but does not change with the surrounding pH. The ratio of light emitted at 520 nm to that at 710 nm thus yields a ratiometric parameter for sensing pH (Fig. 4d, Supplementary Fig. 8a). The 520 nm feature avoids spectral overlap between the 620 nm signal and the broad reference fluorescence near 710 nm. The former initially increases and then decreases with increasing pH, as expected based on the response of the colorimetric filter. The corresponding spectral ratio thus produces the same trend. Because the spectral response is non-monotonic over the full pH range, this metric applies only within the pH interval where the response remains monotonic. Outside this interval, additional RGB or spectral information allows for unambiguous pH determination.

Readout using the spatially resolved approach can be accomplished simply by analysis of RGB values extracted from digital images. The results are as expected, i.e., the ratio in the red channel decreases with increasing pH, while those in the green and blue channels increase (Fig. 4g). The RGB values can be converted into CIE 1931 xy chromaticity coordinates for plotting on a reference chromaticity diagram to provide a standardized visualization of the pH-dependent color changes (Fig. 4j). For the integrated devices, the effective emission lifetime, $t_{1/5}$, defined as the time for the relevant peak emission intensity to decrease to one-fifth of its initial value, is in the range of 15-20 min (Supplementary Fig. 9). These values reflect practical considerations in reliable imaging and ratiometric measurements at meaningful distances and are thus shorter than the >240 min for detectable emission from the bulk CL reaction (Supplementary Figs. 10,11,12). Increasing the initial CL emission intensity by use of the exothermic reaction approach extends the practical sensing range but generally shortens the effective device lifetime. Longer monitoring periods can be readily achieved through sequential deployment of multiple batches of devices. Also, as described previously, direct analysis is possible in the daytime based on intrinsic responses of the colorimetric filters themselves, without the need for CL emission (Supplementary Fig. 13a).

The same analysis approaches apply to sensing of $Hg^{2+}$ and $NO_2^-$. Increasing the concentration of $Hg^{2+}$ progressively attenuates the transmitted CL emission near 520 nm (Fig. 4b), leading to monotonic decreases in both spatially and spectrally resolved ratios (Fig. 4e, Supplementary Fig. 7e,h, Supplementary Fig. 8b). For the former, ratios in the red increase markedly, those in the green change only slightly, and those in the blue decrease in the range of high concentrations $Hg^{2+}$ (Fig. 4h). Corresponding CIE 1931 chromaticity diagrams show the expected $Hg^{2+}$-dependent color shifts (Fig. 4k). Daytime analysis shows similar trends (Supplementary Fig. 13b). For $NO_2^-$ sensing, the CL emission intensities near both 520 nm and 620 nm decrease progressively as the $NO_2^-$ concentration increases from 0 to $10^{-1}$ mol/L (Fig. 4c). Accordingly, the spatially and spectrally resolved ratios decrease monotonically over the range from $10^{-4}$ to $10^{-1}$ mol/L (Fig. 4c,f, Supplementary Fig. 7f,i, Supplementary Fig. 8c). The former yields ratios in the red that increase and those in the green and blue that decrease (Fig. 4i). The corresponding CIE 1931 chromaticity plot also captures the $NO_2^-$-dependent color changes (Fig. 4l). Daytime analysis yields expected results (Supplementary Fig. 13c). The sensing ranges (pH 4-10, $Hg^{2+} \geq$ 1ppm, and $NO_2^- \geq 10^{-4}$ mol/L) encompass environmentally relevant conditions associated with water-quality variation and elevated contamination events.

**Distributed chemical mapping and flow-following sensing**

The envisioned use of fPD devices involves large-scale deployment across aquatic environments, where their buoyant, self-orienting, and light-emitting properties enable broad spatial coverage with accurate readout especially at nighttime. Each device functions as a local, light-emitting chemical sensor capable of detecting environmental parameters such as pH, heavy metals, and nitrite at its deployment location, and along its flow trajectory. Demonstrations under static concentration gradients, dynamic flow conditions, and outdoor pond environments illustrate the potential of this platform across a range of realistic scenarios.

A first example involves $NO_2^-$ monitoring along a meandering channel geometry designed to emulate heterogeneous chemical distributions under controlled static conditions (Fig. 5a; Methods). fPD devices distributed throughout the channel exhibit distinct optical responses according to their local $NO_2^-$ environments (Fig. 5a). The extracted CIE 1931 xy chromaticity coordinates occupy distinct locations in chromaticity space with the reference calibration (Fig. 5b). Analogous experiments using fPD devices for pH sensing yield similar spatially resolved responses (Extended Data Fig. 8).

Additional demonstrations illustrate flow-following measurements under both passive (Extended Data Fig. 9b) and active (Fig. 5c) modes. In meandering-flow experiments (Supplementary Fig. 14) characterized by PIV measurements (Extended Data Fig. 9a), devices released into serpentine channels faithfully follow flow trajectories (Fig. 5d) and report local pH values. Trajectory-derived optical signals can be converted into a normalized color metric, $R^* = \frac{R(t)-R_0}{R_{max}-R_0}$, where $R(t)$, $R_0$ and $R_{max}$ denote the instantaneous, initial and maximum red channel intensities, respectively. Spatial mapping of $R^*$ along the corresponding Lagrangian trajectories yields local chemical conditions that follow underlying flow patterns (Fig. 5e). Corresponding chromaticity coordinates show good agreement with reference calibration values (Extended Data Fig. 9c). Similar behavior occurs in recirculating vortex flow fields (Extended Data Fig. 9d-i).

Experiments in artificial pond environments mimic outdoor conditions, where fPD devices distribute across a shallow water environment containing stones and aquatic plants, with acidic and basic buffer solutions introduced from opposite sides to establish a spatial pH gradient (Fig. 5f left). Imaging with and without ambient lighting reveals position-dependent optical responses (Fig. 5f right). RGB values for the latter case yield CIE 1931 xy chromaticity coordinates and associated local pH values. These discrete data allow for reconstruction into a continuous chemical map using inverse-distance-weighted interpolation (Fig. 5g, Supplementary Fig. 15).

Outdoor field deployments further illustrate device performance in natural aquatic environments near Evanston, Illinois, USA (approximately 100 $m^2$ in area; Extended Data Fig. 10a,c,e). The experiments involve fPD devices deployed simultaneously and imaged remotely during nighttime (Extended Data Fig. 10a,c,e). Quantitative analysis of the resulting images reveals spatially uniform optical responses corresponding to negligible $Hg^{2+}$ concentrations in the tested water body (Extended Data Fig. 10b). Similar field demonstrations for pH and $NO_2^-$ yield measurements (pH $\approx$ 5-6 and $NO_2^-$ $\approx$ 0) consistent with independent reference tests (Extended Data Fig. 10d,f). Together, these results demonstrate the ability of distributed collections of fPD devices to perform remote chemical sensing across static, flow-driven, pond-scale and outdoor environments, establishing a foundation for large-area environmental monitoring through passive deployment and optical imaging.

## Conclusion

The technology introduced here represents a qualitatively distinct method for large-area, monitoring of chemical hazards across natural environments, with complementary features and important advantages relative to other approaches. The scheme utilizes biodegradable materials for light-emitting devices that can be deployed at scale for spatio-temporal mapping, where miniaturized geometries, buoyant device architectures, flow-following transport characteristics, colorimetric sensing layers, and chemiluminescent (CL) sources of light enable wireless operation and accurate, remote assessments without onboard electronics. The results validate the device architectures, CL transduction mechanisms, optical quantification strategies, and image-based readout schemes for operation under both daytime and nighttime conditions. Static-gradient and flow-driven demonstrations within indoor simulated and outdoor natural environments suggest possibilities for practical applications.

Beyond the pH, $Hg^{2+}$, and $NO_2^-$ sensing modes illustrated here, the underlying concepts may apply to other chemical and biological targets given appropriate colorimetric chemistries. Continued advances in chemiluminescent materials, environmentally benign fluorescent reference materials, and handheld, stationary, aerial, or autonomous imaging platforms have the potential to expand both the sensing capabilities and operational lifetimes of these systems. Efforts in these and related directions are of growing importance as we strive to serve as responsible stewards of fragile ecosystems under increasing stresses from anthropogenic activity and extreme natural phenomena.

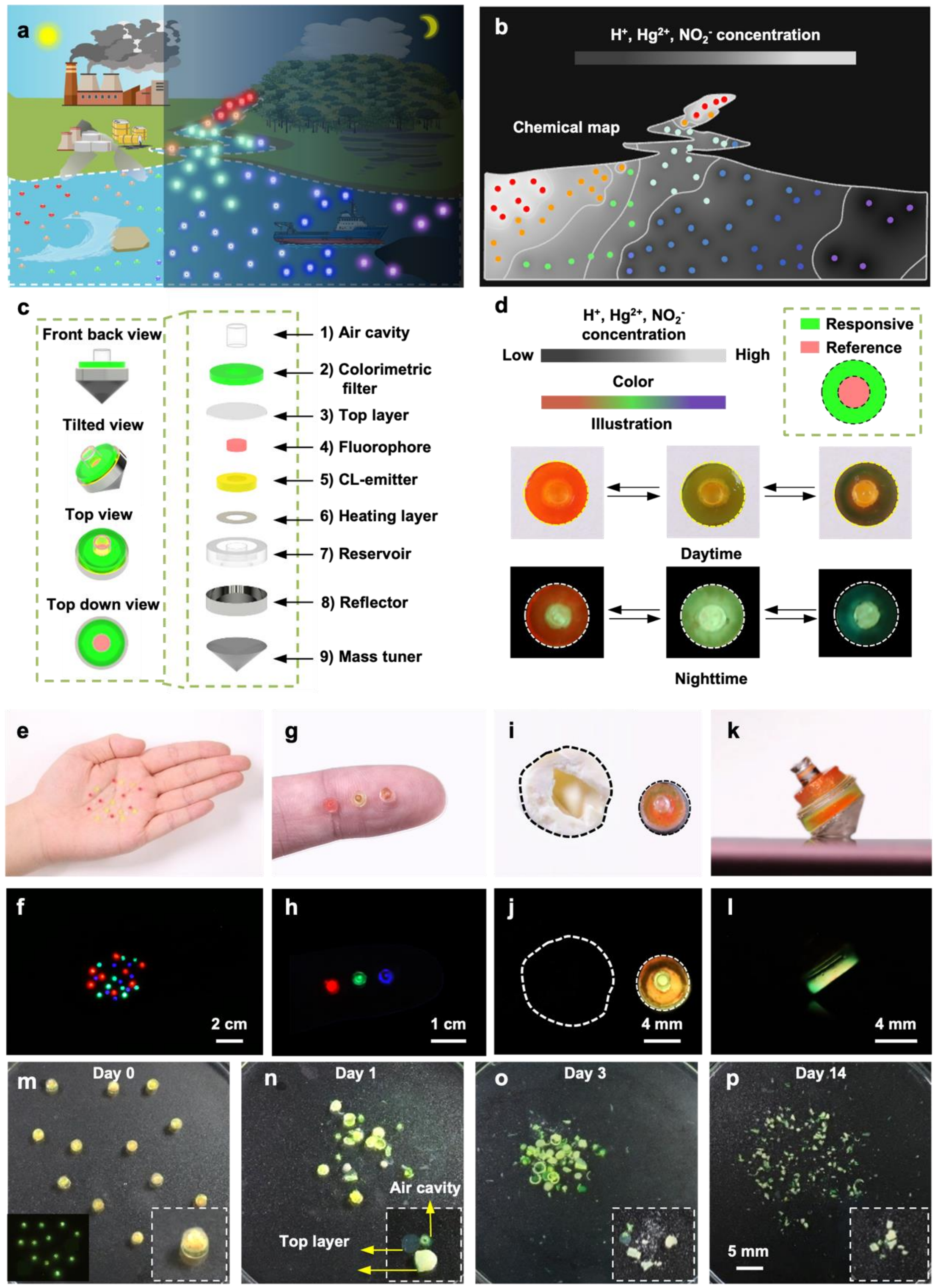

**Figure 1. Functional glowing pixie dust (fPD) devices for distributed environmental sensing by remote, optical readout using active (chemiluminescent light emissive) and passive (ambient-light reflective) modes.** **a,** Illustration of acute environmental events, including anthropogenic activities (toxic airborne release, chemical discharges, oil spills), where widely distributed sensors operate in a passive, ambient light-reflective mode and an CL light emissive mode, optimized for readout during daytime (left) and nighttime (right), respectively. The results offer detailed spatio-temporal information to guide detection and mitigation efforts. **b,** Conceptual illustration showing spatial mapping of representative environmental parameters (pH, heavy metals, and $NO_2^-$). **c,** Exploded-view schematic illustration of the multilayer architecture of a device. **d,** Illustration of reversible

responses to changes in the environmental parameters depicted in **b**. **e, f,** Photographs of representative fPD devices placed on a palm, captured with (**e**) and without (**f**) ambient lighting. **g, h,** Top-view images of a device with (**g**) and without (**h**) ambient lighting. **i, j,** Top-view photographs comparing the size of a device with a dried lotus seed with (**i**) and without (**j**) ambient lighting. **k**, **l**, Side-view photographs of fPD devices without Mg reflectors, showing multilayer architecture. **m-p,** Time-sequenced photographs of accelerated tests of degradation of devices to environmentally benign end products.

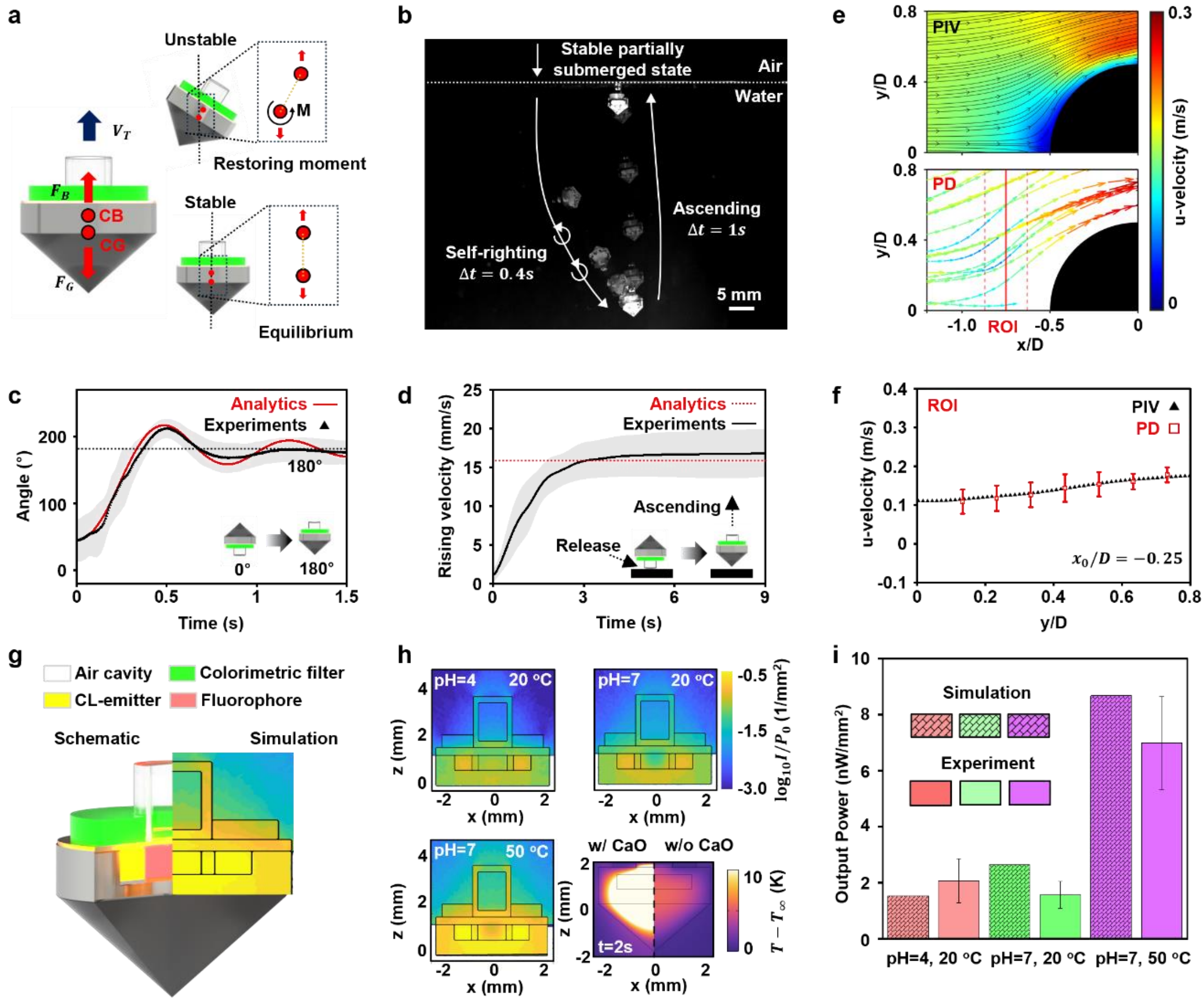

**Figure 2. Hydrodynamic, optical and thermal characteristics. a,** Schematic illustration of a PD highlighting the center of gravity (CG), center of buoyancy (CB), gravitational force ($F_G$), and buoyancy force ($F_B$), together with the self-righting mechanism arising from the offset between CG and CB. **b,** Time-sequenced images showing self-righting and ascent following submersion in water. **c,** Time evolution of the self-righting angle comparing analytical predictions with experimental measurements. **d,** Rising velocity following release from the bottom surface comparing analytical predictions and experimental measurements. **e,** Flow-field characterization by a PD device for representative bluff-body configurations. Time-averaged streamwise velocity fields superimposed by corresponding streamlines of the reference flow via TR-PIV (top), together with reconstructed two-dimensional trajectories colored by local streamwise velocity in the Lagrangian frame of reference (bottom) for the cylindrical bluff body case. **f,** Streamwise velocity profiles extracted upstream of the bluff body, comparing reference PIV measurements with trajectory-derived PD velocities ($x_0/D = -0.25$). **g,** Hybrid schematic illustration of the device architecture and simulation results for the optical field, both shown in cross-section. **h,** Optical and thermal simulations of fPD devices with a reflector. The first three panels show simulated optical output for devices with the colorimetric filter at pH 4 and 20 °C, at pH 7 and 20 °C, and at pH 7 and 50 °C, respectively. The fourth panel shows simulations of temperature increase with exothermic reactions of CaO with water. Modeling treats the CL emission and QD fluorescence as isotropic and homogeneous light sources within their designated domains. The output power corresponds to photons that escape through the top surface. **i,** Experimental measurements and simulations for output power for the three representative conditions shown in **h**: pH 4 at 20 °C, pH 7 at 20 °C, and pH 7 at 50 °C.

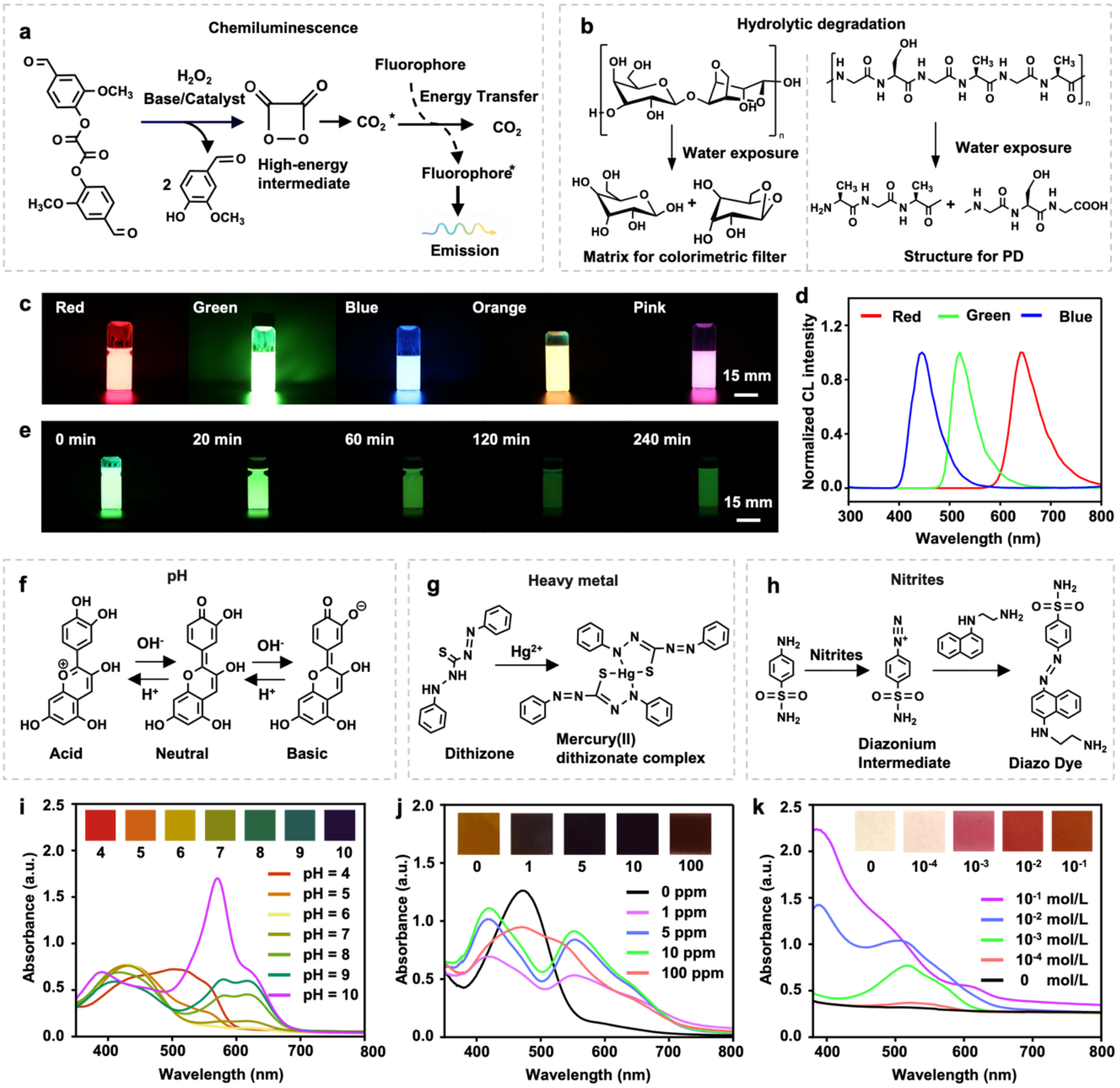

**Figure 3. Chemistries for CL emission and colorimetric sensing. a,** Chemistry of the CL reaction. **b,** Hydrolytic degradation reactions for agarose gel (left; colorimetric filter) and silk fibroin (right; main structural body of the fPD, including the concentric annular reservoir, air cavity, and mass tuner) **c,** Photographs of CL systems that incorporate various fluorescent dyes for emission at wavelengths across the visible range. **d,** Normalized CL emission spectra of a system designed for multicolour emission. **e,** Photographs of a green-emitting CL sample captured at different times. **f, g, h,** Chemistry of reactions associated with colorimetric detection of pH (**f**), heavy-metal concentrations (**g**) and nitric ion concentrations (**h**). **i, j, k,** Ultraviolet-Visible (UV-Vis) absorption spectra of colorimetric filters for different pH levels (**i**), and concentrations of $Hg^{2+}$ (**j**) and $NO_2^-$ (**k**).

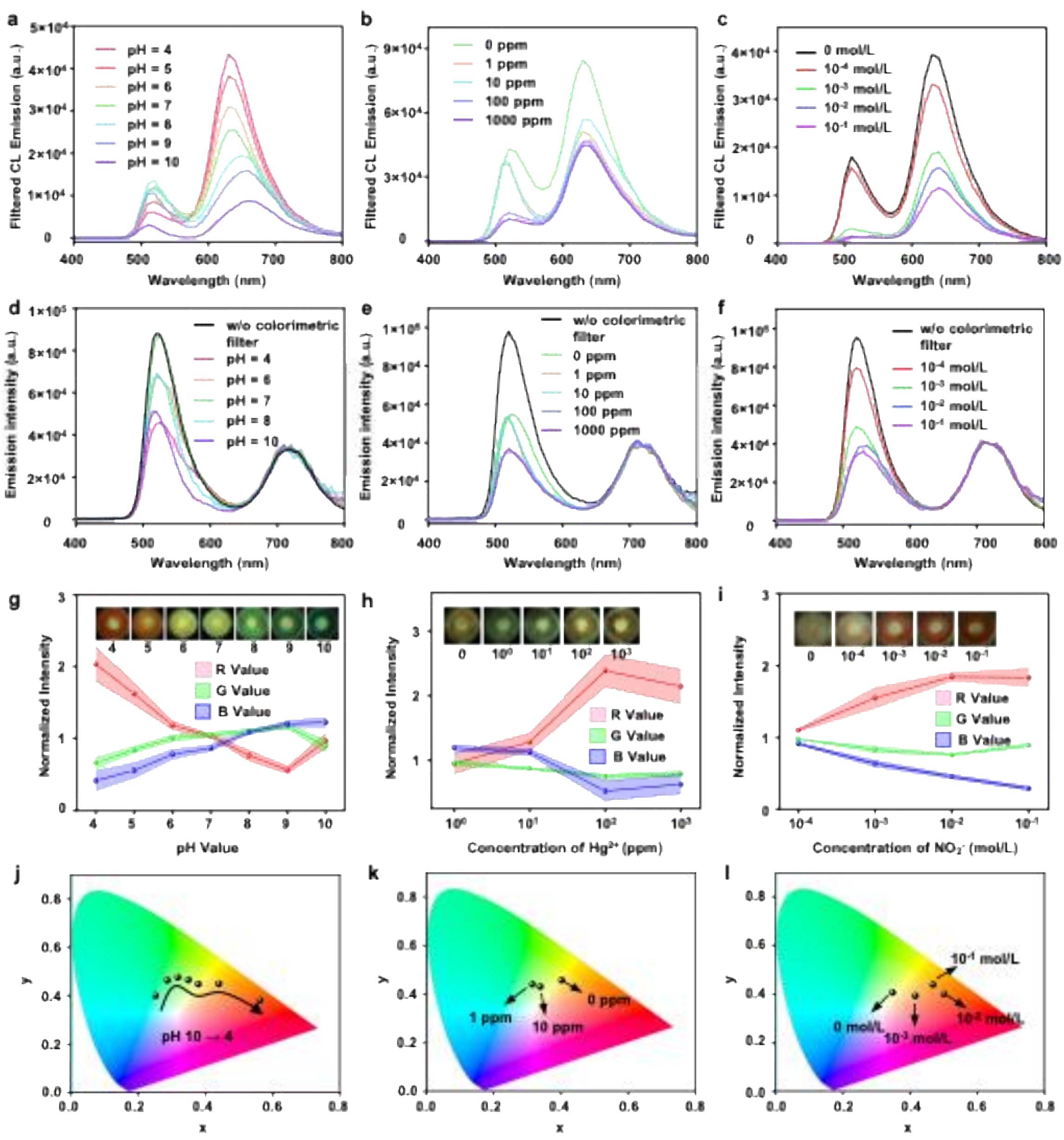

**Figure 4. Quantitative ratiometric sensing in an active mode and emission responses captured by digital imaging. a, b, c,** Spatially resolved spectra of CL emission after passage through a colorimetric filter as a function of pH level (**a**), $Hg^{2+}$ concentration (**b**), and $NO_2^-$ concentration (**c**). **d, e, f,** Spectra of CL emission after passage through a colorimetric filter and normalized by the intensity of CL-excited fluorescence near 710 nm as a function of pH level (**d**), $Hg^{2+}$ concentration (**e**) and $NO_2^-$ concentration (**f**). **g, h, i,** Quantitative analysis of the normalized red, green, and blue channels extracted from images captured without ambient light at different pH levels (**g**), $Hg^{2+}$ concentration (**h**), and $NO_2^-$ concentration (**i**). The circles correspond to mean values and the shaded areas represent the standard deviation (± SD). **j, k, l,** CIE 1931 xy chromaticity coordinates based on RGB values extracted from images captured without ambient light at different pH levels (**j**), $Hg^{2+}$ concentrations (**k**), and $NO_2^-$ concentrations (**l**), and plotted on the reference chromaticity diagram.

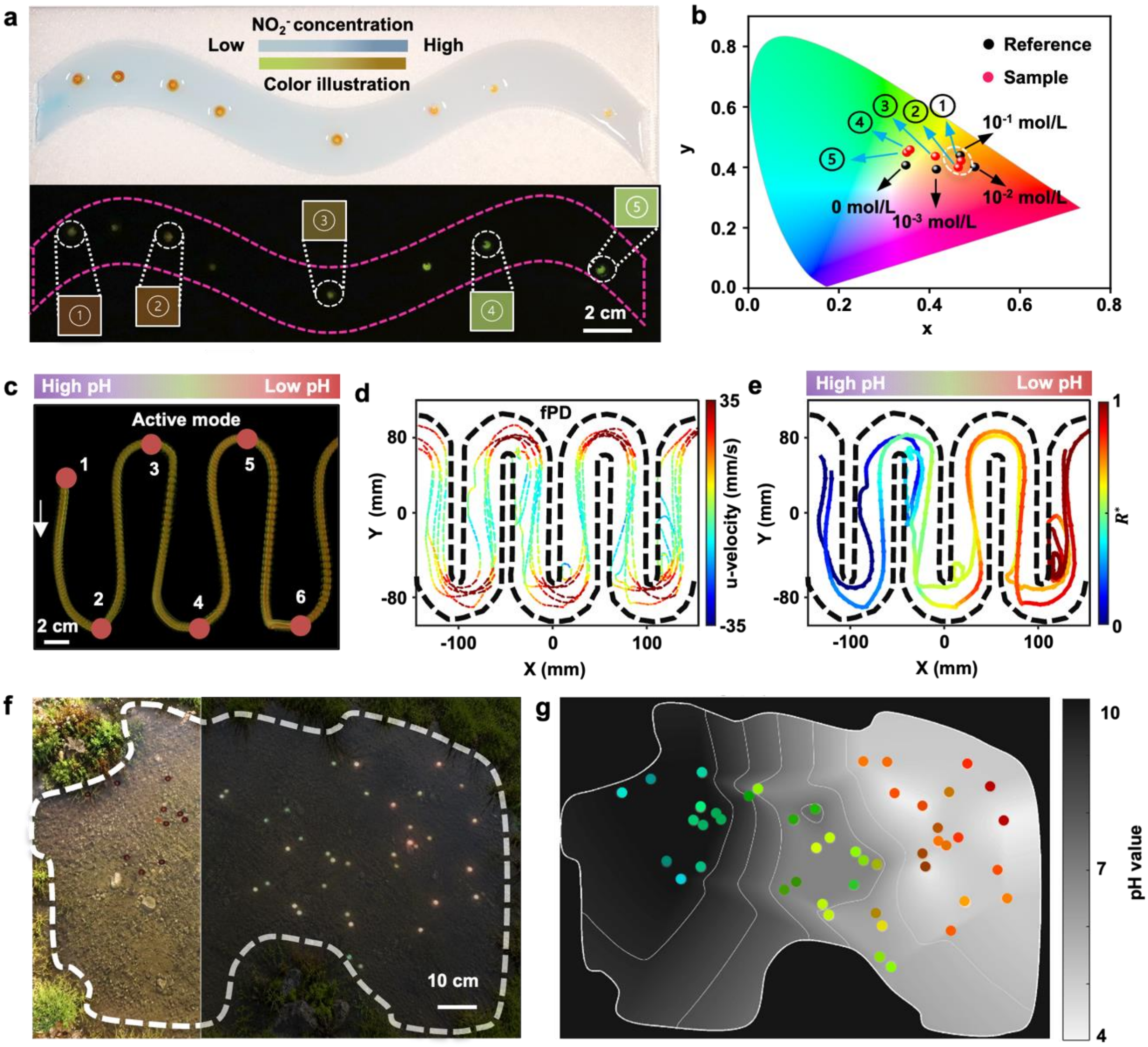

**Figure 5. Functional demonstrations. a,** Spatial mapping of the concentration of $NO_2^-$ using fPD devices in a gel medium. Images with (top) and without (bottom) ambient lighting. **b,** CIE 1931 xy chromaticity coordinates based on RGB values extracted from devices at different positions within the same flow channel and plotted on the reference chromaticity diagram in Fig. 4l. **c,** Distributed pH sensing in serpentine flows using active ($d =$ 13.5 mm**)** mode without ambient lighting. **d,** Reconstructed fPD trajectories colored by local streamwise velocity in the Lagrangian frame of reference. **e,** Reconstructed spatial distribution of the normalized optical response ($R^*$) obtained from distributed fPD trajectories under varying pH conditions. **f,** Field demonstration of 47 fPD devices distributed in a small artificial pond (~1 m²) with separated regions that mimic spatial pH gradients. Images acquired with (**f,** left) and without (**f,** right) ambient lighting. **g,** Spatial pH map of the artificial pond. RGB values extracted from fPD images were converted into CIE 1931 xy coordinates and compared with reference values to determine local pH values in different regions of the pond. The colored circles indicate the corresponding measurement locations, while the grayscale contour map represents the interpolated spatial pH distribution reconstructed from these discrete measurements.

## MATERIALS AND METHODS

### Materials

All reagents, including vanillin, triethylamine, oxalyl chloride, anhydrous dichloromethane (DCM), triacetin, 30% hydrogen peroxide, Coumarin 6, 7-diethylamino-4-methylcoumarin (Coumarin 1), Nile Red, sodium hydroxide (NaOH), agarose, poly(vinyl alcohol) (PVA, Mw 13,000–23,000, 87–89% hydrolyzed), pH indicators, dithizone, 4-aminobenzenesulfonamide, N-(1-naphthyl)ethylenediamine (NED), and quantum dots (CdTe core type quantum dots, COOH functionalized), were purchased from Sigma-Aldrich or Thermo Fisher Scientific and used as received unless otherwise specified.

### Synthesis of CL materials

The process began with dissolving Vanillin (6.4 g, 42.1 mmol) in 100 mL of dichloromethane (DCM) in a 500 mL Erlenmeyer flask while stirring. Adding triethylamine (4.5 g, 44.5 mmol) and stirring the reaction mixture at room temperature for 2 h led to a color change to pale yellow. The next step involved cooling in an ice-water bath (0 °C) and adding oxalyl chloride (2.7 g, 21 mmol) dropwise over a period of 30 min. Removal from the ice bath allowed the reaction to continue at room temperature for another 2 h, prior to quenching by the addition of 100 mL deionized water extraction with DCM (3 × 100 mL). The combined organic layers were washed sequentially with 250 mL of 1.0 M aqueous HCl and then with 200 mL of deionized water. The organic phase was dried over anhydrous magnesium sulfate ($MgSO_4$), filtered, and concentrated under reduced pressure using rotary evaporation to afford a white solid with a mass of 5.5 g, corresponding to a yield of 73% (Supplementary Fig. 2).

### Initiation of the CL reaction

Time-dependent measurements of the intensity of CL emission relied on a reaction mixture formed by first dissolving divanillyl oxalate (21 mg, 7 mg $mL^{-1}$) in 3 mL of triacetin. Thoroughly mixing Coumarin 6 (3 mg, 1 mg $mL^{-1}$) into this solution followed by the addition of 500 μL of 1.0 M aqueous NaOH and then 250 μL of 30% hydrogen peroxide ($H_2O_2$) completed the reaction mixture. Shaking the solution led to bright green CL emission, visible to the eye for more than 120 min. Use of Nile Red and Coumarin 1 yielded red and blue light, respectively. Mixing these dyes led to multi-color emission (Fig 3c, Extended Data Fig. 6a). Orange emission, as used in the fPD devices (1 mL), resulted from mixing the green and red CL solutions at a volume ratio of 25:1.

### Generation of fluorescence

fPD devices based on spectrally selective measurements used CL solution in the outer annular chamber, and a fluorescent reference solution consisting of CdTe quantum dots (QDs, 10 mg/mL in water) in the central chamber. CL emission excited fluorescence from the QDs (Fig. 4d-f).

### Reversible suppression of CL

A freeze-thaw protocol enabled temporary suppression of CL emission during storage and transport, to enable activation at the point of deployment. Placing freshly prepared CL solutions in a dry ice bath (−78 °C) solidified the reaction medium and quenched light emission. Returning the samples to ambient temperature led to a resumption of emission upon thawing (Extended Data Fig. 6d-f). This process also suppressed reaction of the CaO and associated exothermic heating.

**Synthesis of agarose-based pH sensor hydrogels**

The process began with dispersion of agarose (300 mg), poly(vinyl alcohol) (PVA, Mw 13,000–23,000, 87–89% hydrolyzed, 150 mg), and 8.5 mL of a pH-indicator solution based on anthocyanin in 6.5 mL of deionized water. Anthocyanin, extracted from red cabbage, undergoes a pH-dependent structural transformation and corresponding change in absorption properties, as the basis of the chromophore in this solution. Specifically, under acidic conditions, the flavylium cation form is dominant, producing a red appearance. As the pH increases, deprotonation induces conversion to quinoidal and related species, leading to progressive changes in the absorption spectrum and a visible color shift toward green and purple. Heating the mixture to 70 °C under constant stirring led to complete dissolution of the agarose and PVA to yield a homogeneous solution. Casting the solution between two glass slides separated by a 0.5 mm spacer, followed by cooling to room temperature formed a uniform hydrogel film with a thickness of 0.5 mm. The films were stored in the dark at room temperature prior to use.

**Synthesis of agarose-based mercury(II) sensor hydrogels**

The procedure for forming mercury(II)-responsive hydrogels began with dissolving agarose (150 mg) and poly(vinyl alcohol) (PVA, 150 mg) in 15 mL of deionized water at 90 °C to yield a clear and homogeneous solution, subsequently maintained at 70 °C to prevent gelation. Separately, dissolving dithizone (75 mg) in 15 mL of acetone yielded a 20 mM stock solution for colorimetric sensing. Vortex mixing the agarose/PVA solution (15 mL) with 2 mL of a solution formed by dissolving dithizone (75 mg) in 15 mL of acetone yielded a uniform, orange solution. Dithizone undergoes coordination interactions with divalent metal ions, particularly $Hg^{2+}$. Specifically, the sulfur- and nitrogen-containing donor sites chelate $Hg^{2+}$, forming a metal–ligand complex that alters electron delocalization within the conjugated chromophore. This interaction changes both the intensity and spectral distribution of the absorption bands. An increasing concentration of $Hg^{2+}$ progressively modifies the absorption spectrum, including splitting of the original band into two features, thereby providing the basis for concentration-dependent optical discrimination. Formation of a mercury(II)-responsive film from this solution followed the same procedures as those described above for pH sensing, with storage in the dark at room temperature.

**Synthesis of agarose-based nitrite sensor hydrogels**

Preparation of nitrite-responsive hydrogels relied on the Griess reaction system. Dissolving agarose (150 mg) and poly(vinyl alcohol) (PVA, 150 mg) in buffer solution (10 mL, pH 3.0) at 90 °C obtained a clear homogeneous solution. The next step involved dissolving N-(1-naphthyl)ethylenediamine dihydrochloride (NED, 10 mg) and sulfanilic acid (10 mg) in warm buffer (~5 mL), adding the resulting solution to the hot agarose/PVA solution, and mixing thoroughly. Sensing in this case follows from a diazotization–coupling reaction that generates an azo dye chromophore in situ. In the presence of nitrite, an aromatic amine precursor first forms a diazonium intermediate, which subsequently couples with a second aromatic component to produce a conjugated azo dye. Formation of this chromophore increases the absorption in the visible range and gives rise to a concentration-dependent color change. As the nitrite concentration increases, the absorbance rises markedly, especially over 400–600 nm, with a pronounced increase near ~620 nm, thereby providing the basis for quantitative optical analysis. Formation of a nitrite-responsive film from this solution followed the same procedures as those described above for the pH and mercury(II) cases, also with storage in the dark at room temperature.

## Fabrication of a bioresorbable reservoir structure

A previously established protocol produced regenerated silk fibroin (RSF) from Bombyx mori silk cocoons. Briefly, cutting the cocoons into small pieces and boiling them in 0.02 M sodium carbonate ($Na_2CO_3$; Sigma-Aldrich) solution for 30 min removed the sericin coating. Drying the degummed fibers and then dissolving them in a 9.3 M lithium bromide (LiBr; Sigma-Aldrich) solution at 60 °C for 1 h yielded an RSF solution. Dialysis against deionized water (DIW) for 3 consecutive days, using a membrane with a molecular weight cutoff (MWCO) of 3.5 kDa, removed residual chaotropic ions. Centrifugation at 11,400 g for 20 min at 4 °C purified the solution by removing aggregated particulates. Freezing the purified RSF solution overnight at −80 °C, and subsequent lyophilization (Labconco; −80 °C, 0.006 bar) produced a dried RSF solid. Grinding the dried RSF with a ball mill produced a fine powder[31]. Thermal pressing of the RSF powder produced a dense, solid silk block. Placing 5 g of the powder in a 1-inch square pellet die and pressing at 90 °C under approximately 14 MPa (Carver, Auto Series Plus) for 30 min yielded a silk block with a thickness of approximately 5 mm. CNC milling (Roland) machined the block into a cylindrical cup geometry. A dental polishing tool subsequently refined the tip of each cup to a tapered profile. Formation of the concentric annular reservoir, air cavity and mass tuner structures followed similar procedures. The same thermal pressing process, using 0.3 g of RSF powder, yielded a top-layer film with a thickness of approximately 0.3 mm.

## Assembly of the devices

Electron-beam evaporation of a layer of Mg (300 nm thick) onto a film of silk fibroin (10-µm thick) formed the reflector structure. Integration involved cutting this film into narrow strips, and bonding it onto the sidewalls and bottom surface of the reservoir. Placing a ring-shaped CaO structure in the reservoir, followed by bonding of the mass tuner, concentric annular reservoir, and top layer by softening the contacting surfaces with a small amount of silk solution prepared the main body of the device for bonding of the air cavity onto the center of the top layer. Punches (4 mm and 1.5 mm diameters) patterned the colorimetric filter into annular ring geometries, mechanically fitted to the air cavity. Injecting the CL solution into the outer annular chamber of the concentric annular reservoir and the QD fluorescence reference solution into the central reservoir using a syringe pump (NE-300, New Era Pump Systems) equipped with a 34G needle completed the process (Fig. 1c). The small puncture formed by the 34G needle did not lead to noticeable leakage during handling or freezing.

## Environmental degradation test

Chemical dissolution tests of degradation involved immersion of fPD devices in Ajisawa's reagent[32,33] ($CaCl_2$:ethanol:water = 1:2:8 molar ratio) at 70 °C. Removing samples from this solution once per day allowed collection of photographs to document the stages of structural decay. Under these conditions, within the first day the device exhibited structural collapse, separation of individual components, and complete dissolution of the Mg reflector. By day 14, only small remnants of the silk reservoir/cup and fragmented debris remained (Fig. 1m-p).

## Characterization of absorption spectra of colorimetric filters

A standard spectrophotometer (UV–Vis, LAMBDA 1050) captured the absorption spectra of the filters over a wavelength range of 300–800 nm. Measurements involved films exposed to analyte solutions with different pH values or concentrations (Fig. 3i-k). A corresponding unexposed film served as a reference for background correction.

## Characterization of the spectral properties of light produced by CL reactions

A multi-mode plate reader (BioTek Synergy Neo2) recorded the spectra of light emitted by CL reactions, at the default setting of 20 °C unless otherwise specified. Measurements at 20, 30, 40, and 50 °C, established by the reader platform, defined the temperature dependence across a relevant range (Extended Data Fig. 5c). Evaluations at 0 and 10 °C involved first cooling the samples by placing them on dry ice and then allowing them to gradually warm by exposure to room temperature while monitoring with a thermal camera, to allow measurements immediately as temperature approached 0 or 10 °C.

## Characterization of spectra associated with the active mode of sensing

Measurements used pre-mixed CL solutions containing fluorophores with addition of 50 μL of 1 M NaOH and 50 μL of 30% $H_2O_2$ to initiate the reaction, cast in wells of a 24-well plate (Fig. 3d). Placing colorimetric filters over the wells allowed for characterization of transmitted light (Supplementary Fig. 6). For fully integrated devices incorporating inorganic fluorophores, spectral measurements involved the assembled devices, with comparisons to corresponding spectra obtained through the sensing layer (Fig. 4a-f).

## Characterization of output optical power with the active mode of sensing

Measurements used a photometer equipped with a Thorlabs S120VC sensor immediately after injection of 40 μL of CL solution into devices with and without reflector structures under otherwise identical conditions (Extended Data Fig. 4a-d).

## Spatially resolved ratiometric analysis

One variant of spatially resolved ratiometric analysis uses the ratio of the intensity of CL emission transmitted through the annular colorimetric filter to that from the central region, evaluated across a range of wavelengths. The other approach, suitable when the digital imaging approach cannot resolve the central and annular regions of a device, uses instead the ratio of CL emission from a device that includes a colorimetric filter to a spatially adjacent one that does not, in an otherwise similar manner (Fig. 4a-c).

## Spectrally resolved ratiometric analysis

Spectrally resolved ratiometric analysis used the ratio of CL emission transmitted through the colorimetric filter to the CL-excited fluorescence (Fig. 4d-f). The time-dependent evolution of the CL emission spectrum during the reaction must be considered. Measurements indicate that the absolute CL intensity decreases over time, but the overall spectral shape remains largely unchanged. As a result, the reaction time mainly affects the overall signal intensity rather than the spectral profile. Because QD fluorescence follows from the same CL reaction, the ratiometric analysis eliminates effects of variations in CL emission intensity during the measurement period.

**Image capture and color analysis**

Placing colorimetric filters above the CL solutions enabled capture of digital images (Canon EOS 6D camera) without ambient lighting at different times after CL activation. RGB values from annular and central regions of the device followed from analysis using commercial software (Image J) (Fig 4g-i, Supplementary Fig. 10). The following equation defines normalized channel values, $N_i = {}^{i}/_{(R+G+B)}, i = R, G\ or\ B.$ where i represents the R, G, or B channel value. For RGB-based spectrally resolved ratiometric analysis, the normalized ratio is $R_i = {}^{N_{i,filter}}/_{N_{i,ref}}$, $(i = R, G, B)$, where $N_{i,filter}$ and $N_{i,ref}$ denote values from the annular and central regions, respectively.

**Temperature measurement**

Studies of heating associated with reactions of CaO used dry ice placed in a large Petri dish, with a smaller Petri dish containing water positioned above it. The system equilibrated to a water temperature of ~10 °C. Injection of the CL solution into the fPD device, followed by placement into the water immediately preceded the capture of infrared thermal videos using a FLIR A655sc IR camera positioned ~20 cm above the device (Extended Data Fig. 5a,b). The temperature scale was 0 to 35 °C for all measurements.

**Hydrodynamic characterization**

Studies of hydrodynamic self-righting and ascent used in a water-filled acrylic chamber (15 × 15 × 20 $cm^3$) and a high-speed camera (HT-2000M, Emergent) equipped with a 25 mm F/1.4 lens (LM25HC, Kowa) at a resolution of 2048×1088 pixels. Illumination with light emitting diodes (LEDs) allowed recordings at 200 Hz for more than 10 s, yielding 2,000 frames per trial for devices released from a height of 18 cm above the water surface. Quantitative measurements of trajectory and orientation followed from two-dimensional particle tracking velocimetry (2D-PTV; Fig. 2a-d). To minimize external disturbances, a custom enclosure enabled force-free release through dissolution of a temporary water-soluble adhesive, thereby ensuring reproducible initial conditions. Image processing and trajectory analysis followed previously reported procedures.[18]

Particle image velocimetry (PIV) defined the instantaneous hydrodynamic flow fields. In these experiments, devices released from a height of 50 cm remained within a continuous laser sheet (PSU-W-LED, Opto Engine LLC). The PIVlab toolbox in MATLAB with a two-pass FFT-based window deformation algorithm served as the basis for analyzing image sequences of 500 frames acquired at 100 Hz. The first pass used a 64 × 64 pixel interrogation window, followed by a final pass with a 16 × 16 pixel window and 50% overlap. A spatial calibration factor of 0.150 mm/pixel yielded a final vector spacing of 1.20 mm. These conditions produced a valid detection probability (VDP) of approximately 98% (Extended Data Fig. 3a).

**Spatial mapping of $NO_2^-$**

Experiments used a 3D-printed serpentine channel with a width of 25 mm and a height of 30 mm (Fig. 5a), filled with ~300 mL of 3 wt% agarose gel, as a matrix to suppress rapid diffusion and maintain spatial concentration gradients. The experiments began with gradual introduction of a10 mL aliquot of 0.1 M $NO_2^-$ standard solution mixed with 10 mg of water-soluble blue dye for visualization into the left side of the channel. Deployment of fPD devices along the channel for optical readout occurred shortly after setting of the gel and stabilization of the concentration field (Fig. 5a).

**Bluff body flow mapping**

Experiments employed a circulating water tunnel with a test section of 0.3 × 0.7 × 0.25 $m^3$ under a uniform free-stream velocity of 0.17 m/s, corresponding to a Reynolds number of $8.5 \times 10^3$ and $1.36 \times 10^4$. Bluff bodies formed by 3D-printing consisted of a circular cylinder (diameter: 50 mm) and a rectangular bar (cross-section of 30 × 80 $mm^2$), each with a vertical height of 160 mm and mounted vertically at the center of the test section. TR-PIV within a laser-illuminated plane (PSU-W-LED, Opto Engine LLC) characterized the Eulerian flow fields. A high-speed CMOS camera (EoSens 1.1 CXP2, Mikrotron) equipped with a 50 mm F/1.4 lens (Kowa LM50HC) recorded image sequences at a resolution of 1280 × 864 pixels and a sampling frequency of 500 Hz for 13 s. Velocity fields followed from recursive cross-correlation analysis using 64 × 64 pixel interrogation windows in the first pass and 16 × 16 pixel final windows with 50% overlap in final pass. The resulting vector spacings were 1.28 mm and 1.34 mm for the cylindrical and rectangular bluff bodies, respectively. These conditions yielded VDPs of approximately 99%.

Trajectory measurements used aerial imaging acquired with a drone (Matrice 4T, DJI) positioned 1 m above the water surface. Placing the PD devices at the water surface under streamwise flow approximately 30 cm upstream of the bluff bodies using tweezers, preceded their gentle release to minimize external disturbances. Images acquired at a resolution of 3840 × 2160 pixels and a sampling frequency of 30 Hz for 10 s enabled characterization of the device kinematics. Streamwise velocities followed directly from tracked trajectories using the 2D-PTV framework described above. Flow traceability was evaluated by comparing velocity profiles derived from device trajectories with those obtained from PIV measurements. Conversion of Lagrangian trajectory data into Eulerian profiles used a moving-window spatial averaging approach. Agreement between the two datasets confirms that the PD devices accurately follow mean flow structures and reliably reconstruct spatial velocity variations (Fig. 2e,f, Extended Data Fig. 3c,d).

**Meandering flow mapping**

Meandering-flow experiments used a 3D-printed serpentine open channel with a width of 40 mm and a height of 30 mm mounted within the test section of a circulating water tunnel (Supplementary Fig. 14). The water depth was 20 mm. Dissolution of 15 g of anhydrous sodium carbonate ($Na_2CO_3$, DAEJUNG) set the background solution to pH ~9.5 while an acidic reservoir solution consisted of 20 g of citric acid ($C_6H_8O_7$, DAEJUNG) dissolved in 1 L of water. Time-resolved particle image velocimetry (TR-PIV) within an LED-illuminated region characterized the Eulerian flow fields. A high-speed CMOS camera (IRON 250M-RCG, KAYA) equipped with a 25 mm F/1.4 lens (Kowa LM25HC) recorded image sequences at a resolution of 2096 × 1024 pixels and a sampling frequency of 100 Hz for 50 s. Velocity fields followed from recursive cross-correlation analysis using 64 × 64 pixel interrogation windows in the first pass and 16 × 16 pixel final windows with 50% overlap in final pass, yielding valid detection probabilities (VDPs) of approximately 98%. Measurements of fPD trajectories used images acquired with a mobile camera (A2411, Apple) at a resolution of 2160 × 3840 pixels and a sampling frequency of 10 Hz for more than 30 s. Device trajectories and local velocities followed from the 2D-PTV framework described above. RGB values, corresponding to pH levels, extracted from the sensing region using ImageJ yielded normalized values according to $R^* = \frac{R(t)-R_0}{R_{max}-R_0}$, where $R(t)$ is the red value at time $t$, $R_{max}$ is the maximum red value along each trajectory, and $R_0$ is the initial red value. In active sensing mode, contrast-enhanced images generated using ImageJ enabled extraction of red-channel intensity variations under low-light conditions (Fig. 5c-e, Extended Data Fig. 9a-c).

**Vortex flow mapping**

Vortex-flow experiments used a water-filled acrylic chamber (15 × 15 × 20 $cm^3$) with a water depth of 10 cm. A magnetic stir bar (27 mm diameter) rotated counterclockwise at 100 rpm, at the center of the chamber to generate a stable vortex flow. Dissolution of 0.2 g of anhydrous sodium carbonate ($Na_2CO_3$, DAEJUNG) adjusted the background solution to ~9.5. Introduction of 5 g of citric acid powder ($C_6H_8O_7$, DAEJUNG) near the stir bar led to mixing-driven dissolution and a dynamically evolving pH gradient. TR-PIV within a laser-illuminated plane (PSU-W-LED, Opto Engine LLC) characterized the Eulerian flow fields. A high-speed CMOS camera (EoSens 1.1 CXP2, Mikrotron) equipped with a 35 mm F/1.4 lens (Kowa LM35HC) recorded image sequences at a resolution of 1280 × 864 pixels and a sampling frequency of 50 Hz for 40 s. Velocity fields followed from recursive cross-correlation analysis using 64 × 64 pixel interrogation windows in the first pass and 16 × 16 pixel final windows with 50% overlap in final pass. These conditions yielded VDPs of approximately 99%. Trajectory measurements used images acquired with a mobile camera (A2411, Apple) at a resolution of 2160 × 3840 pixels and a sampling frequency of 30 Hz for more than 5 min. Velocity magnitudes followed directly from tracked trajectories using the 2D-PTV framework described above. RGB values extracted from the sensing region using ImageJ enabled quantitative analysis of dynamic color evolution during vortex transport. Ground-truth pH measurements relied on a multiparameter meter (HI 98194, Hanna instruments; Extended Data Fig.9).

**Indoor pond demonstration for large-area pH mapping.**

The indoor artificial pond consisted of a 1.2 m × 1.2 m container with pebbles, rocks, and aquatic plants distributed throughout to mimic an outdoor aquatic environment. Water at pH 7 filled the container to a depth of 2–3 cm. Deployments included 47 fPD devices, including seven reference devices. A spatial pH gradient resulted from the gradual introduction of a buffer solution at pH 10 from the left side of the pond, and a buffer solution at pH 4 from the right side. Image capture occurred after approximately 5 min without ambient light. RGB values extracted from each device yielded corresponding CIE 1931 xy chromaticity coordinates. Comparisons of these coordinates to a calibration map (Fig. 4j). allowed for assignment of pH values corresponding to the nearest calibration point in CIE chromaticity space (Supplementary Fig. 15). A continuous map of pH followed from the application of inverse-distance weighted interpolation based on pH values at the spatial positions of the deployed devices.

**Acknowledgements** This work was supported by the Querrey Simpson Institute for Bioelectronics at Northwestern University. J-T. K. acknowledges funding from National Research Foundation of Korea (NRF) grant funded by the Korea government (MSIT)(RS-2024-00342270) and (MSIT and MOE) (No. RS-2025-17492968).

**Author contributions** Z.H., D.Z. and J.A.R. initiated and conceived the functional pixie dust concept. Z.H., D.Z., J.-T.K. and J.A.R. designed the studies and analysed the results. Z.H., D.Z., J.C., J.Z., Q.Z., L.Z., S.X. and R.C. fabricated and characterized the functional pixie dust devices. J.K., D.S., U.J., H.K., S.L., U.-S.B. and J.-T.K. performed hydrodynamic simulations. H.Z., K.Z. and Y.H. performed optical and diffusion simulations. X.Y. and Y.H. performed thermal simulations. Z.H. and C.P. developed and synthesized the chemiluminescent materials. Z.H., D.Z., Q.Z., L.Z., L.Y., Q.W. and Z.X. conducted the fluorescence characterization. T.K. G.B. and W.M. fabricated the biodegradable silk structures, D.Z. and J.Z. integrated them into devices. Z.H. and D.Z. designed and fabricated the colorimetric filters. D.Z. and J.C. developed fluorophore layer and analysed the corresponding data. D.Z., J.Z. and S.X. developed heating layer and performed thermal experiment and analysed the corresponding data. D.Z., J.C., J.Z. and T.K. performed the biodegradation tests. D.Z., Z.H., J.K. J.C. and J.Z. performed the indoor and outdoor experiments and analysed the corresponding data. Z.H., D.Z., J.K., H.Z., J.C., C.P., Q.Z., S.X., X.Y., D.S., Y.H., J.-T.K. and J.A.R. discussed and interpreted the data. Z.H., D.Z., J.-T.K. and J.A.R. prepared the figures and wrote the paper. In addition, H.Z., X.Y. and Y.H. assisted with the preparation of figures and text. Z.H., D.Z., J.-T.K. and J.A.R. revised the paper. Z.H., D.Z., J.K., H.Z., J.C. and C.P. contributed equally to this work.

**Competing interests** The authors declare no competing interests.